\documentclass{PoS}

\title{An update in monopole condensation in two-flavour Adjoint QCD}
\ShortTitle{Monopole condensation in $N_f=2$ aQCD: an update}

\author{\speaker{Giuseppe Lacagnina}\\ Dipartimento di Fisica and
        INFN, Pisa, Italy\\ E-mail: \email{lacagnina@df.unipi.it}}

\author{Guido Cossu\\ Scuola Normale Superiore and INFN, Pisa, Italy\\
E-mail: \email {g.cossu@sns.it}} \author{Massimo D'Elia\\ Dipartimento
di Fisica and INFN, Genova, Italy\\ E-mail:
\email{Massimo.Delia@ge.infn.it} } \author{Adriano Di Giacomo\\
Dipartimento di Fisica and INFN, Pisa, Italy\\ E-mail:
\email{digiaco@df.unipi.it}} \author{Claudio Pica\\ Brookhaven
National Laboratory, Physics Department, Upton, NY 11973-5000, USA\\ E-mail: \email{pica@bnl.gov}}

\abstract{Two distinct phase transitions occur at different
  temperatures in QCD with adjoint fermions (aQCD): deconfinement and
  chiral symmetry restoration. We study monopole condensation by
  inspecting the expectation value of an operator which creates a
  monopole: such a quantity is expected to be an order parameter for
  the deconfinement transition as in the case of fermions in the
  fundamental representation. Our results are consistent with a first
  order deconfinement transition. Furthermore, the magnetic order
  parameter is found to be unaffected by the chiral transition.}

\abstract{QCD with fermions in the adjoint representation (aQCD) is a
model for which a deconfinement and a chiral phase transition take
place at different temperatures. In this work, we present a study of
the deconfinement transition in the dual superconductor picture based
on the evaluation of an operator which carries magnetic charge. The
expectation value of this operator signals monopole condensation and
is an order parameter for deconfinement as in the case of fermions in
the fundamental representation. We find a sharp first order
deconfinement transition. We also study the effects of the chiral
transition on the monopole order parameter and find them negligible.}

\FullConference{The XXV International Symposium on Lattice Field Theory\\
		 July 30-4 August 2007\\
		 Regensburg, Germany}

\begin{document}

\section{Introduction}

\noindent Ordinary QCD, for which quarks are in the fundamental
representation of $SU(3)$, shows two phase transitions, deconfinement
and chiral symmetry restoration, which according to numerical
simulations take place at the same temperature. The coincidence of the
two transitions makes it very difficult to study the degrees of
freedom which are relevant for each transition separately. On the
other hand, QCD with fermions in the adjoint representation of $SU(3)$
(aQCD), is a model for which the two phase transitions seemingly
happen at distinct temperatures \cite{karsch, engels}. Furthermore,
contrary to the case of standard QCD, quarks in the adjoint
representation do not explicitly break the $Z(3)$ symmetry of the
action, and the Polyakov loop is a good order parameter for
deconfinement.

In this work, we study the deconfinement phase transition in the Dual
Superconductor Picture (DSP), in which confinement follows from dual
superconductivity of the QCD vacuum, realized as condensation of
magnetic charges \cite{thooft}. To study monopole condensation, we
construct a magnetically charged operator whose expectation value
vanishes exactly in the deconfined phase and becomes nonzero in the
confined phase \cite{pisa1,pisa2, pisa3, pisa4}, thus defining an
order parameter. Our main goal is to take advantage of the features of
aQCD to investigate the relation between dual superconductivity and
the dynamics of chiral symmetry breaking. In particular, we study the
behaviour of the magnetic order parameter in the proximity of the
chiral phase transition.

The authors of \cite{karsch, engels} have performed simulations of aQCD
with $N_f=2$ staggered quarks, and found a deconfinement and a chiral
phase transition at different temperatures, with $\beta_{\rm
dec}<\beta_{\rm chiral}$. The chiral transition has been further
investigated in \cite{engels}, where the authors made an extended
analysis to determine its order. Their results from the magnetic
equation of state indicated a second order chiral transition in the
$3d \ O(2)$ universality class in the zero quark mass limit.

The outline is as follows. In Section \ref{aQCDmono} we briefly
describe aQCD and the monopole condensation order parameter. We
summarize simulation details in Section \ref{simulations}. We discuss
our results in Section \ref{results} and draw conclusions Section
\ref{conclusions}.

\section{aQCD and monopole condensation}
\label{aQCDmono}

\subsection{aQCD}

\noindent Quark fields in the adjoint representation of $SU(3)$ can be
written as $3\times 3$ hermitian traceless matrices
\begin{equation}
Q(x) = Q^a(x)\lambda_a
\end{equation}
in terms of Gell-Mann's $\lambda$ matrices. For the fermionic sector
of the action one therefore needs to use the 8-dimensional, real
representation of the link variables:
\begin{equation}
U^{ab}_{(8)} = \frac{1}{2}{\rm Tr}\left
(\lambda^aU_{(3)}\lambda^bU_{(3)}^{\dagger}\right )
\label{eq_u8}
\end{equation}
Including the pure gauge sector, in which links are in the usual
3-dimensional representation, the full action reads:
\begin{equation}
S = S_G[U_{(3)}] + \sum_{x,y} {\bar Q}(x)M\left (U_{(8)}\right
)_{x,y}Q(y)
\end{equation}
where $M$ is the fermionic matrix. The Polyakov loop is defined by
\begin{equation}
L_{3} \equiv \langle \frac{1}{3L_s^3}|\sum_{{\vec x}}{\rm
  Tr}\prod_{x_0=1}^{L_t}U_0^{(3)}(x_0,{\vec x})|\rangle
\end{equation}
and is an order parameter for the spontaneous breaking of the center
symmetry. The Polyakov loop is related to the free energy of an
isolated static quark in a gluonic bath at temperature $T$: $L_3
\propto e^{-F/T}$ \cite{svetitsky}. This result justifies its use as
an order parameter for the deconfinement transition: the free energy
is infinite in the confined phase implying $L_3=0$, while it is finite
in the deconfined phase ($L_3\neq 0$).

\subsection{Monopole condensation}

\noindent A possible order parameter for the deconfinement phase
transition is given by the vacuum expectation value of a magnetically
charged operator \cite{pisa1,pisa2, pisa3, pisa4}. This operator adds
a magnetic monopole to a given gauge field configuration: a non
vanishing expectation value is the signature of monopole condensation
and of the Higgs breaking of the underlying magnetic
symmetry. On the other hand, in the deconfined phase, where the
symmetry is restored, the {\it vev} of the magnetic operator drops to
zero. The evaluation of this order parameter involves a few steps. One
starts by fixing the gauge with a procedure known as Abelian
Projection \cite{abelian}. However, as shown in \cite{pisa2}, the
particular gauge choice does not affect the behaviour of
the order parameter. In practice, we can update the system without an
Abelian Projection, which is equivalent to choosing a different random
gauge at each step \cite{pisa3}. Next, for each configuration, the
values of the action are evaluated in presence and in absence of a
monopole field insertion in the temporal plaquettes of a given time
slice \cite{pisa1}. Then, the expression for the order parameter is:
\begin{equation}
\langle \mu \rangle = \frac{1}{Z}\int [dU] e^{-{\widetilde
S}}=\frac{{\widetilde Z}}{Z}
\end{equation}
where ${\widetilde S}$ is modified by the presence of the monopole
field. A much easier quantity to evaluate is however
\begin{equation}
\rho = \frac{\partial}{\partial\beta} \ln \langle \mu \rangle =
\langle S\rangle_S - \langle\widetilde{S}\rangle_{\widetilde{S}}
\end{equation}
which is expected to have a large negative drop at the transition
point. Close to the transition ($\beta\simeq\beta_{\rm dec}$), a
scaling behaviour of the type
\begin{equation}
\rho \simeq L^{1/\nu}f( L^{1/\nu}(\beta_c-\beta))
\end{equation}
is expected (for some function $f$), with $\nu=1/3$ for a first order
transition (scaling with spatial volume).

\section{Simulation details}
\label{simulations} 

\noindent We simulate two flavours of staggered quarks on two
lattices, with sizes of $L_s^3\times L_t=12^3\times 4, 16^3\times 4$,
and bare quark masses of $am_q=0.01, 0.04$ for several values of
$\beta$ in the range $3.0-7.0$. Since the evaluation of $\rho$
requires two simulations for each value of $\beta$, we use the exact
RHMC algorithm \cite{rhmc} in presence of the monopole insertion and
the $\Phi$ algorithm \cite{duane} otherwise. Typical MD trajectories
have a length of $N_{MD}\delta t=0.5$, and an integration step $\delta
t = 0.02 - 0.005$ depending on the mass. We employ the Conjugate
Gradient to invert the fermion matrix. We also implement $C^*$
boundary conditions as they become necessary when simulating a
monopole insertion \cite{cstar}. The simulations run on the ApeMille
machine in Pisa and the ApeNEXT facility in Rome.

\section{Results}
\label{results}

\begin{figure}
\includegraphics[width=0.85\textwidth]{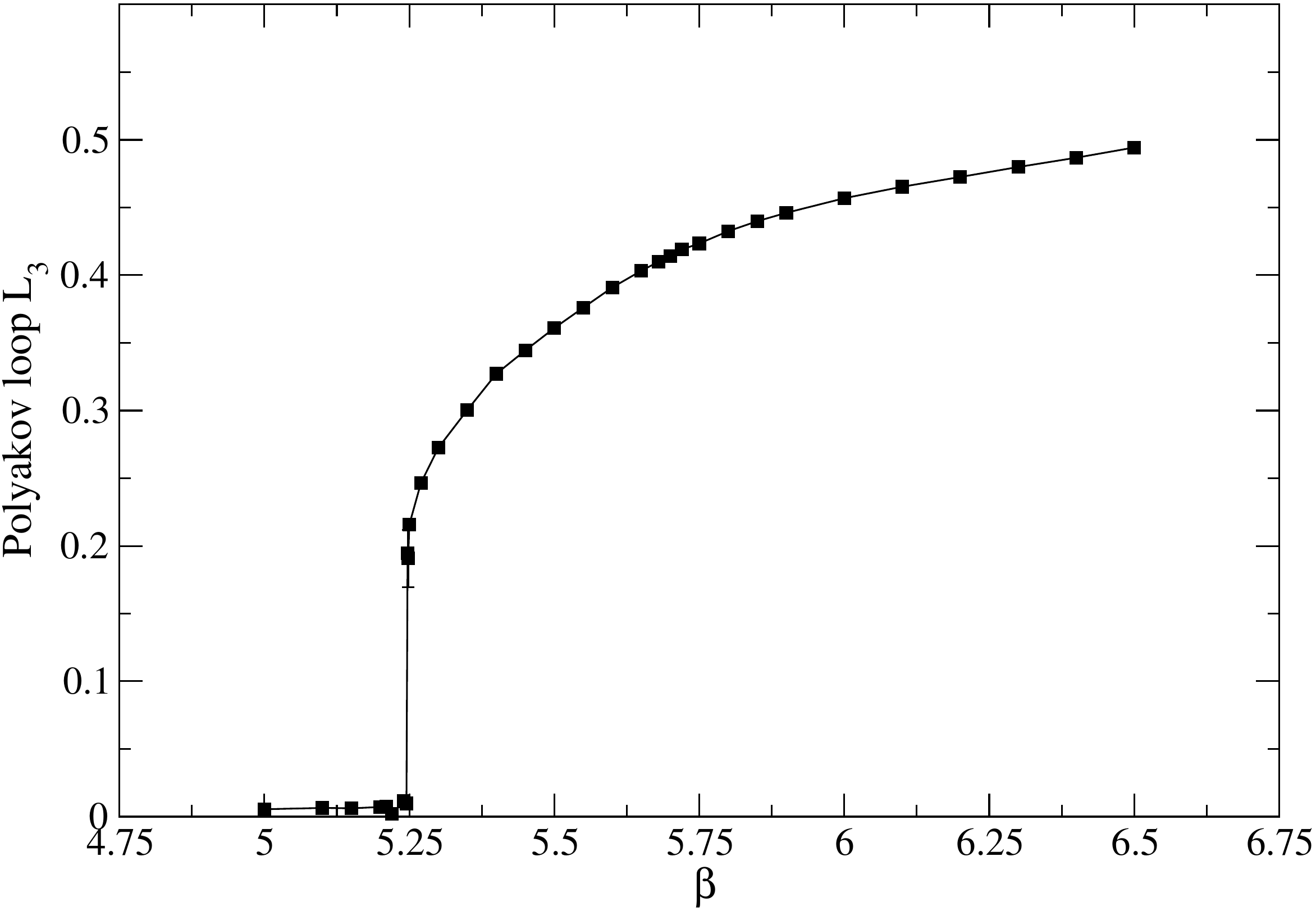}
\caption{The Polyakov loop, with $am=0.01$ and $16^3\times 4$
lattice. Lines are left to guide the eye.}
\label{polyakov}
\end{figure}

\begin{figure}
\includegraphics[width=0.85\textwidth]{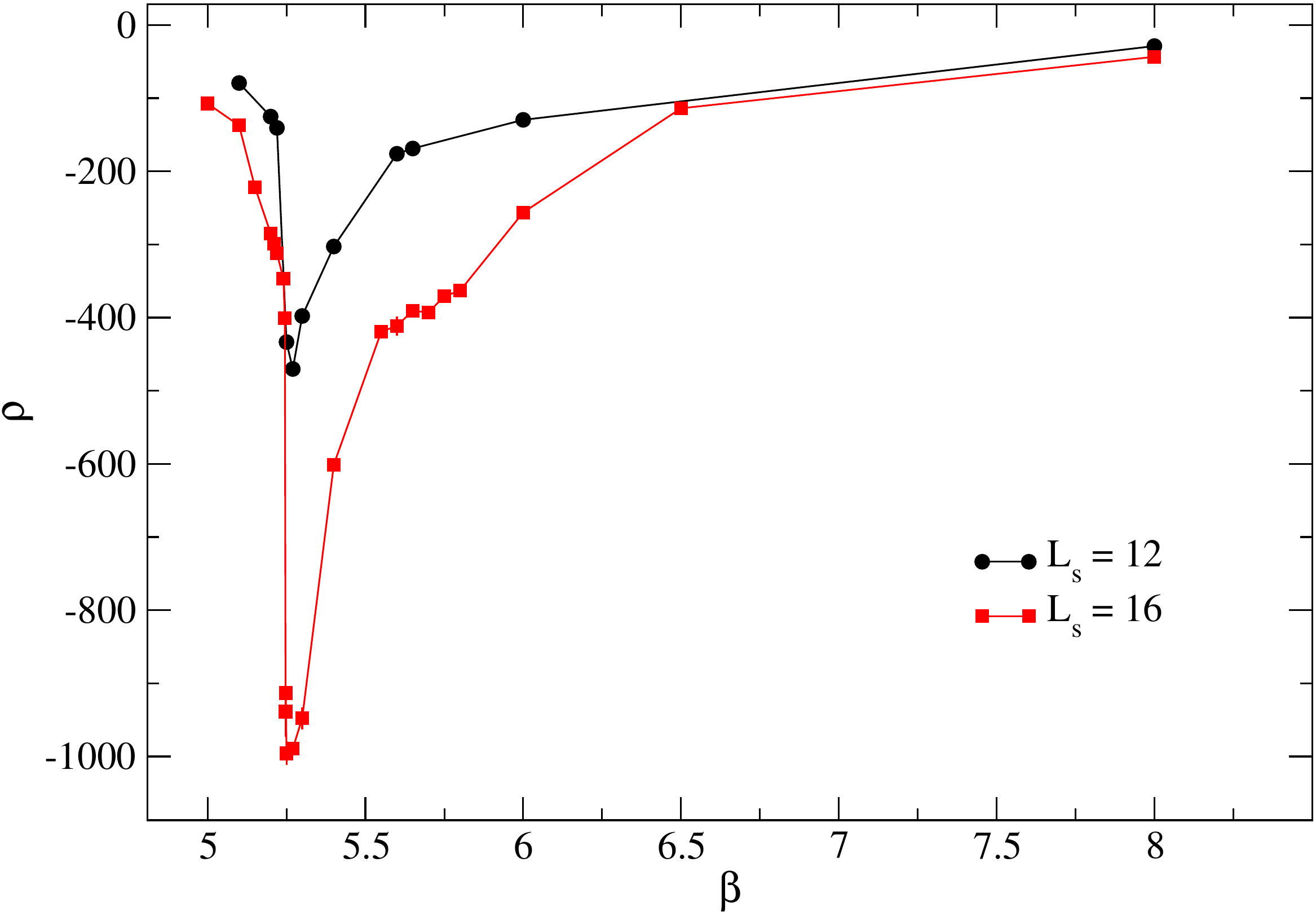}
\caption{The $\rho$ parameter, with $am=0.01$, $L_t=4$, for two different
  spatial volumes.}
\label{plot_rho}
\end{figure}

\begin{figure}
\includegraphics[width=0.85\textwidth]{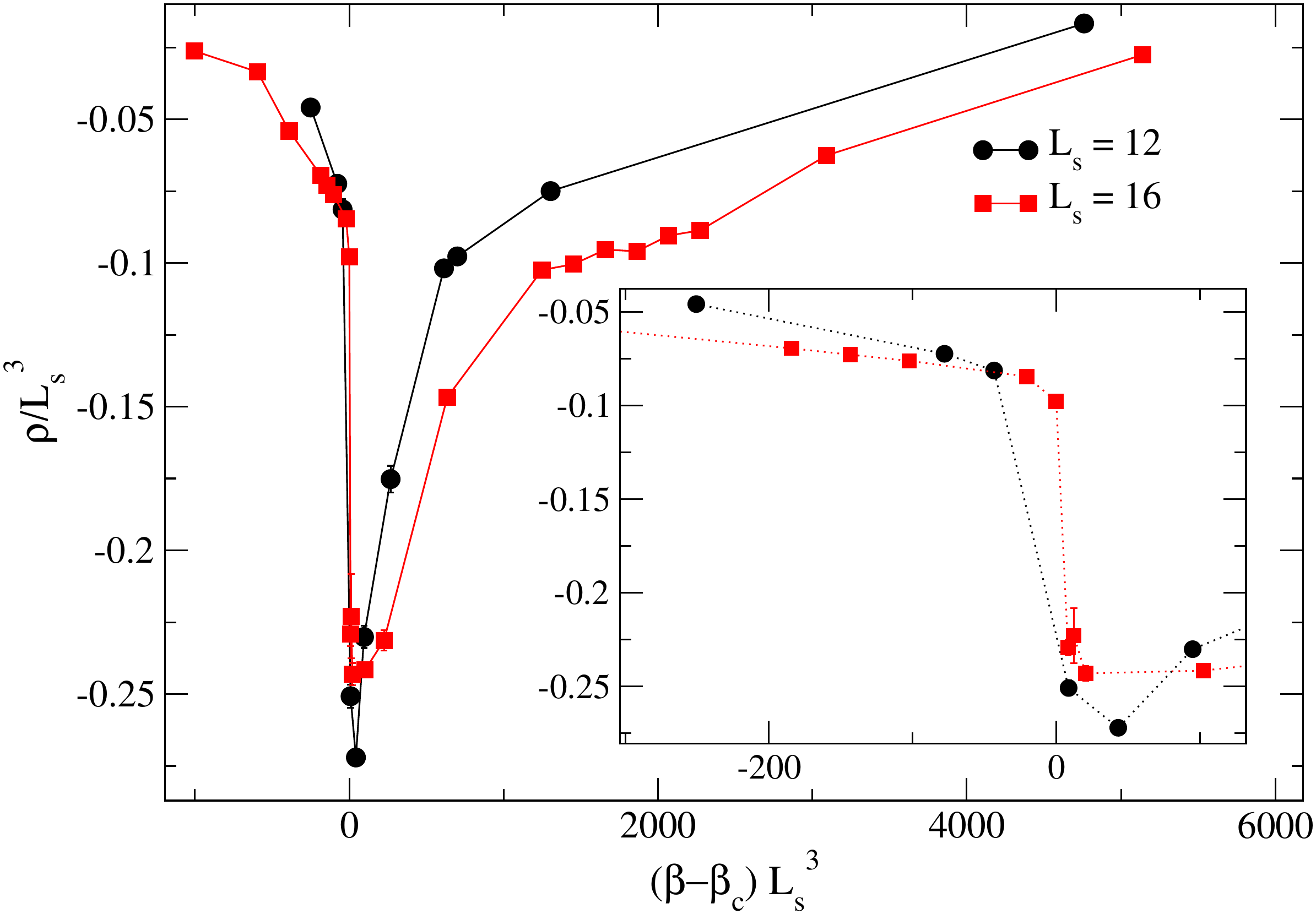}
\caption{Scaling of the $\rho$ parameter, $am=0.01$,
$L_t=4$. $\beta_c=5.25$, estimated from the Polyakov loop at $L_s=16$.}
\label{rho_scaling}
\end{figure}

\begin{figure}
\includegraphics[width=0.85\textwidth]{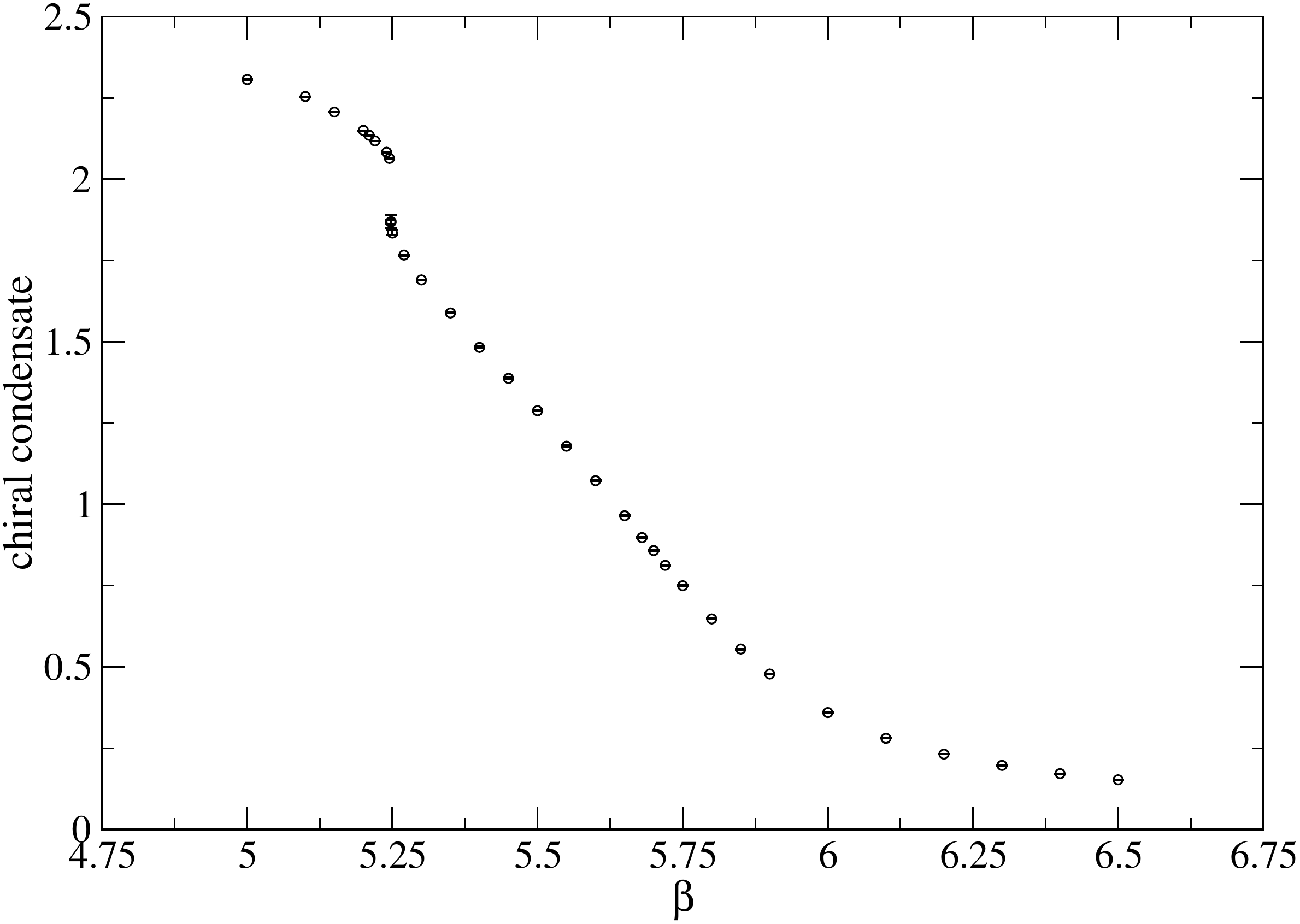}
\caption{Chiral condensate, with $am=0.01$, $16^3\times 4$ lattice. A
clear jump is also visible at the the deconfinement phase transition.}
\label{chicond}
\end{figure}

\begin{figure}
\includegraphics[width=0.85\textwidth]{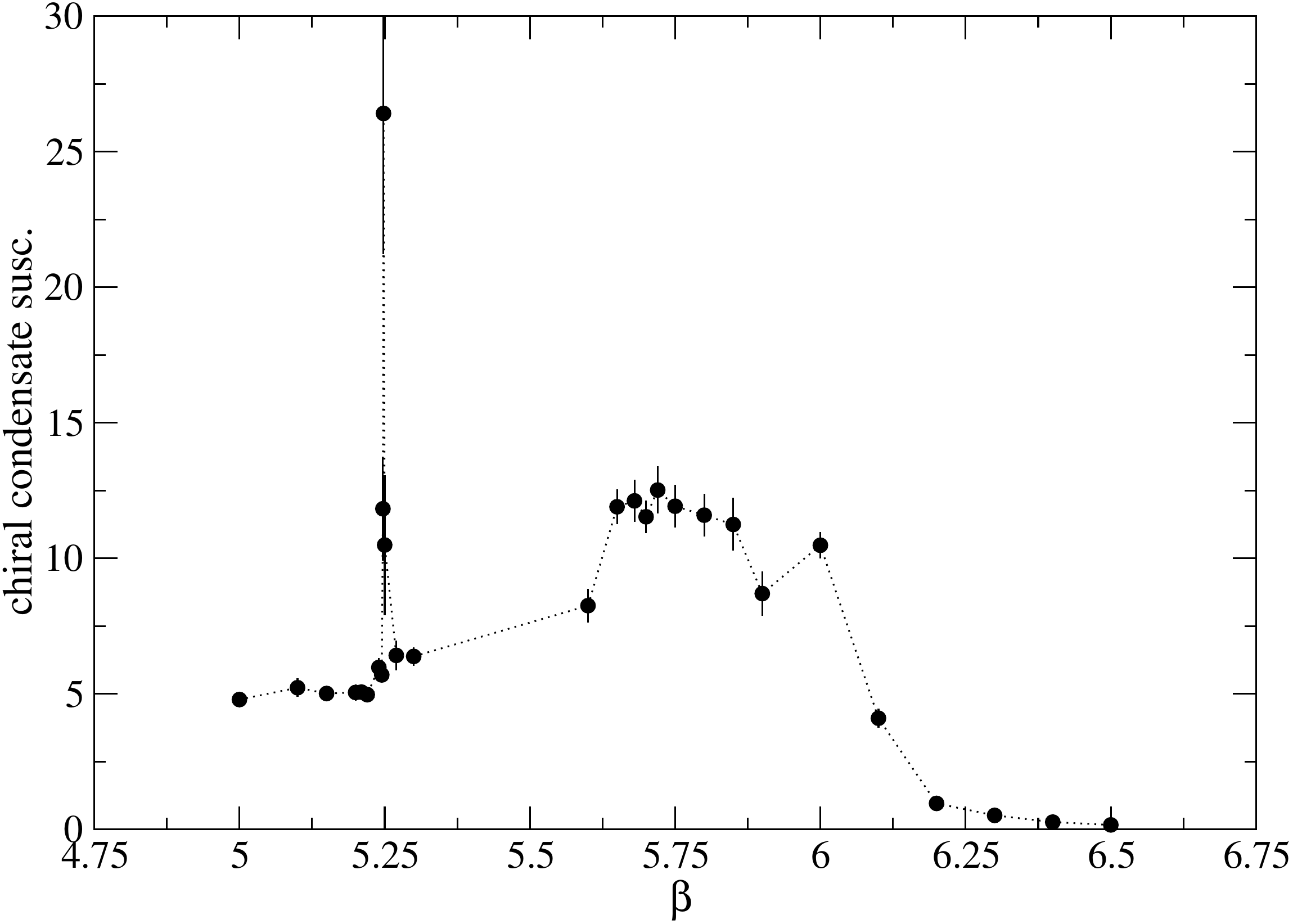}
\caption{Susceptibility of the chiral condensate, with $am=0.01$,
$16^3\times 4$ lattice.}
\label{chisusc}
\end{figure}

\noindent For each configuration, we evaluate the plaquettes, the modified plaquette ---  in a different simulation --- to calculate the $\rho$ parameter, the Polyakov loop, the chiral condensate and its susceptibility. The Polyakov Loop, has the behaviour of a sharp first order transition
for $\beta\simeq 5.25$ (see Fig. \ref{polyakov}). We take this value
of the pseudocritical $\beta$ as an estimate for $\beta_{\rm dec}$ in
our finite size scaling analysis.
We evaluate the $\rho$ parameter and study its scaling behaviour. The
expected negative peak is found at values of $\beta$ which are
compatible with the discontinuity of the Polyakov loop
(Fig. \ref{plot_rho}).
By finite size scaling analysis, we find that $\rho$ has the scaling
properties of an order parameter for a first order transition for both
values of the quark mass (see Fig. \ref{rho_scaling} and \cite{tucson}
for $am=0.04$).
Our main interest is that of finding possible effects of the chiral
transition on the magnetic order parameter. To do so, we first perform
a rough localization of the chiral transition by inspection of the
chiral condensate and its susceptibility (Figs. \ref{chicond},
\ref{chisusc}).
Our results for the light mass $am=0.01$ are compatible with with a
chiral transition around $\beta_{\rm chiral}=5.8$, in agreement with
\cite{karsch, engels}. For both the simulated quark masses, we find
that $\rho$ does not change significantly in the vicinity of the
chiral transition. We conclude that condensation of monopoles,
associated to confinement, is a property only of the gauge sector of
the theory and is not affected by the chiral transition.

\section{Conclusions}
\label{conclusions}

\noindent We presented the results of our study on the effects of the
chiral transition on monopole condensation for QCD with two flavours
of staggered fermions in the adjoint representation. The appeal of
this model is in the fact that its chiral and deconfinement phase
transitions happen at distinct temperatures, making it possible to
study the effects of one transition on the order parameter of the
other. Within the framework of the Dual Superconductor Picture, we
study the {\it vev} of a magnetically charged operator which signals
monopole condensation and is expected to be an order parameter for
deconfinement. Our analysis indicates a first order deconfinement
transition, and the magnetic order parameter is found to be unaffected
by the chiral transition. This result gives further evidence to the
idea that the DSP mechanism of confinement is independent of the
presence of fermions \cite{pisa4}.

The work of C.P. has been supported in part by contract DE-AC02-98CH1-886 with the U.S. Department of Energy.

\end{document}